\documentclass[conference]{IEEEtran}
\IEEEoverridecommandlockouts

\usepackage{cite}
\usepackage{enumitem}
\usepackage{amsmath,amssymb,amsfonts}
\usepackage{algorithmic}
\usepackage{graphicx}
\usepackage{textcomp}
\usepackage{xcolor}
\usepackage{forest}
\usetikzlibrary{trees,positioning,shapes,shadows,arrows.meta}
\usepackage{tikz}
\usepackage{adjustbox} 

\makeatletter
\newcommand{\linebreakand}{%
  \end{@IEEEauthorhalign}
  \hfill\mbox{}\par
  \mbox{}\hfill\begin{@IEEEauthorhalign}
}
\makeatother

\begin{document}

\title{Seeing is Deceiving: Exploitation of Visual Pathways in Multi-Modal Language Models}

\author{
    \IEEEauthorblockN{
        Pete Janowczyk\textsuperscript{\#},
        Linda Laurier\textsuperscript{+}, 
        Ave Giulietta\textsuperscript{\dag},
        Arlo Octavia\textsuperscript{\ddag},
        Meade Cleti\textsuperscript{*, \S}
    }
    \IEEEauthorblockA{
        \textsuperscript{\#}University of Wisconsin - Madison, USA
    }
    \IEEEauthorblockA{
        \textsuperscript{+}Hampton College, USA
    }
    \IEEEauthorblockA{
        \textsuperscript{\dag}Texas A\&M University, USA
    }
    \IEEEauthorblockA{
        \textsuperscript{\ddag}Liberty University, USA
    }
    \IEEEauthorblockA{
        \textsuperscript{\S}Arizona State University, USA
    }
    \IEEEauthorblockA{
        *Corresponding Email: mcleti@asu.edu
    }
}

\maketitle


\begin{abstract}
Multi-Modal Language Models (MLLMs) have transformed artificial intelligence by combining visual and text data, making applications like image captioning, visual question answering, and multi-modal content creation possible. This ability to understand and work with complex information has made MLLMs useful in areas such as healthcare, autonomous systems, and digital content. However, integrating multiple types of data also creates security risks. Attackers can manipulate either the visual or text inputs, or both, to make the model produce unintended or even harmful responses. This paper reviews how visual inputs in MLLMs can be exploited by various attack strategies. We break down these attacks into categories: simple visual tweaks and cross-modal manipulations, as well as advanced strategies like VLATTACK, HADES, and Collaborative Multimodal Adversarial Attack (Co-Attack). These attacks can mislead even the most robust models while looking nearly identical to the original visuals, making them hard to detect. We also discuss the broader security risks, including threats to privacy and safety in important applications. To counter these risks, we review current defense methods like the SmoothVLM framework, pixel-wise randomization, and MirrorCheck, looking at their strengths and limitations. We also discuss new methods to make MLLMs more secure, including adaptive defenses, better evaluation tools, and security approaches that protect both visual and text data. By bringing together recent developments and identifying key areas for improvement, this review aims to support the creation of more secure and reliable multi-modal AI systems for real-world use.
\end{abstract}

\section{Introduction}

\subsection{Overview of MLLMs and Security Implications}

Multi-Modal Language Models (MLLMs) have emerged as a significant advancement in artificial intelligence, integrating both visual and textual data to enhance understanding and interaction capabilities \cite{radford2021learning, liu2023llava, li2024surveying}. MLLMs uses sophisticated architectures that combine vision encoders with powerful language models, enabling applications such as image captioning, visual question answering, and multi-modal content generation \cite{ye2023mplugowl, bai2023qwen, sun_qin_peng_2024}. The fusion of different data modalities allows MLLMs to perform tasks that require a nuanced interpretation of both visual and linguistic information, thereby expanding their utility across various domains including healthcare, autonomous systems, and digital content creation \cite{liu2024safety}.

The integration of multiple modalities in MLLMs broadens their attack surface, making them vulnerable to adversarial attacks across both visual and textual data types \cite{shayegani2023jailbreak}. Unlike single-modality models, MLLMs can be manipulated through either modality, enabling more complex attacks that alter outputs in unintended ways \cite{xu2024cross}. This cross-modal vulnerability allows adversarial inputs in one pathway to affect the other, often bypassing traditional security measures designed for single-modal systems \cite{peng2024securing}.

\subsection{Significance of Visual Attack Vectors}

Visual attack vectors pose a serious risk to MLLMs by subtly altering images or videos to mislead the models into generating inaccurate or harmful text, despite changes often being imperceptible to humans \cite{chen2020fawa, li2024images, yin2023vlattack}. These attacks exploit the complex ways MLLMs process visual and textual information together, weakening data interpretation and security \cite{zhou2024adversarial}. This vulnerability is especially critical in high-stakes areas like autonomous driving, where such manipulations could jeopardize safety, and healthcare, where they could lead to incorrect diagnoses or treatment advice \cite{fu2024pg, li2024survey}. These exploits undermine model trustworthiness and endanger both user privacy and system security \cite{cui2023robustness}.

\subsection{Paper Organization}

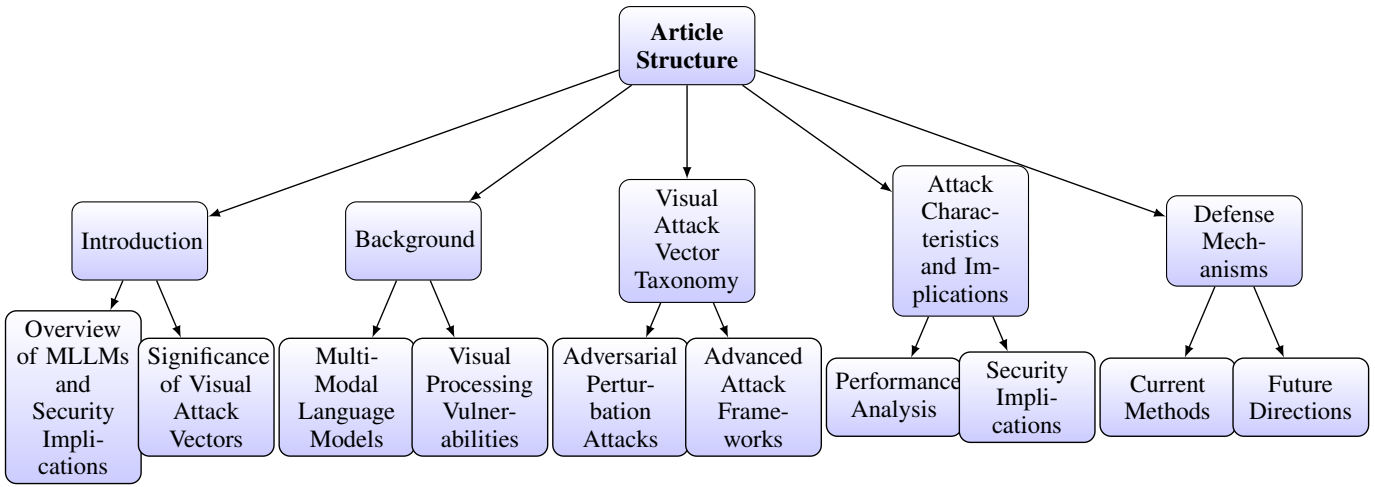
\begin{figure*}[ht]
    \centering
    \begin{adjustbox}{width=\textwidth}
    \begin{tikzpicture}[
        level 1/.style={sibling distance=35mm, level distance=25mm},
        level 2/.style={sibling distance=17mm, level distance=20mm},
        level 3/.style={sibling distance=2mm, level distance=20mm},
        every node/.style={
            rectangle, 
            draw, 
            rounded corners, 
            align=center, 
            top color=white, 
            bottom color=blue!20,
            text width=1.5cm,
            minimum height=1cm,
            font=\small
        },
        edge from parent/.style={
            draw, 
            -latex,
            line width=0.5pt
        }
    ]
        \node {\textbf{Article Structure}}
            child { node {Introduction}
                child { node {Overview of MLLMs\\and Security Implications} }
                child { node {Significance of Visual Attack Vectors} }
            }
            child { node {Background}
                child { node {Multi-Modal Language Models}
                }
                child { node {Visual Processing Vulnerabilities}
                }
            }
            child { node {Visual Attack Vector Taxonomy}
                child { node {Adversarial Perturbation Attacks}
                }
                child { node {Advanced Attack Frameworks}
                }
            }
            child { node {Attack Characteristics and Implications}
                child { node {Performance Analysis}
                }
                child { node {Security Implications} }
            }
            child { node {Defense Mechanisms}
                child { node {Current Methods}
                }
                child { node {Future Directions}
                }
            };
    \end{tikzpicture}
    \end{adjustbox}
    \caption{Hierarchical Structure of the Review Article}
    \label{fig:article_structure}
\end{figure*}

This article reviews how visual pathways in MLLMs can be exploited. After the introduction, Section \ref{sec:background} covers basic information on MLLMs, including how they’re built, how different parts work together, and key security issues. Section \ref{sec:taxonomy} categorizes different types of visual attacks used against MLLMs. In Section \ref{sec:analysis}, we examine how these attacks affect model performance and security. Section \ref{sec:defense} outlines current defenses and considers ways to strengthen MLLMs against these attacks. Section \ref{sec:conclusion} wraps up with main takeaways and possible directions for future research. See \textbf{Fig \ref{fig:article_structure}}.

\section{Background}\label{sec:background}

\subsection{Multi-Modal Language Models}

\subsubsection{Architecture Overview}

Multi-modal language models (MLLMs) integrate visual and textual data through architectures that combine vision encoders and language models, facilitating comprehensive understanding across modalities. A common approach involves using a pre-trained vision encoder to extract visual features, which are then aligned with textual representations within a language model. For instance, the mPLUG-Owl model employs a visual knowledge module alongside a language model, enabling effective multi-modal learning \cite{ye2023mplugowl}.

Another strategy is to utilize cross-attention mechanisms to fuse visual and textual information. The Qwen-VL model integrates a vision transformer with a large language model through a vision-language adapter, employing cross-attention to align visual and linguistic data \cite{bai2023qwen}. Similarly, the LLaVA model connects a pre-trained language model with a visual encoder using a simple linear layer, allowing for efficient processing of image-text instructions \cite{liu2023llava}.

Some architectures incorporate intermediate fusion techniques, where visual and textual inputs are processed separately before integration. The Idefics2 model uses a SigLIP vision encoder and a Mistral-7B language model, with a Perceiver pooling layer to map image features to a fixed number of visual tokens, facilitating robust text encoding \cite{laurenccon2024matters}.

\subsubsection{Vision-Language Integration Mechanisms}

Multi-modal language models (MLLMs) integrate visual and textual data to enhance contextual understanding and output generation. A primary method for this integration is cross-modal attention, which aligns visual and textual elements by computing attention weights that highlight relevant features across modalities. This approach enables the model to focus on pertinent visual regions when processing text and vice versa, facilitating effective information fusion. For instance, the Multi-Modality Cross Attention Network (MCAN) employs cross-modal attention to jointly model intra- and inter-modality relationships between image regions and sentence words, improving image-sentence matching tasks \cite{wei2020multi}.

Another integration strategy is feature fusion, which combines features from different modalities into a unified representation. This can be achieved through various techniques, such as concatenation, summation, or more sophisticated methods like attention-based aggregation. For example, the Compound Tokens method introduces channel fusion for vision-language representation learning, where visual and textual features are combined along the channel dimension to create compound tokens, enhancing alignment between modalities \cite{aladago2022compound}.

Additionally, models like Perceiver and Perceiver IO utilize asymmetric attention mechanisms to process inputs from multiple modalities, including images and text, without modality-specific components. These models use a latent bottleneck to distill inputs, allowing them to handle large-scale, heterogeneous data effectively \cite{jaegle2021perceiver}. The Language Grounded QFormer integrates a Query Transformer (QFormer) with language grounding techniques to align visual and textual representations efficiently, improving vision-language understanding tasks \cite{choraria2023language}. Similarly, the VisionGPT framework consolidates state-of-the-art foundation models to facilitate vision-language understanding, enabling applications such as text-conditioned image understanding and visual question answering \cite{kelly2024visiongpt}.

These integration mechanisms are crucial for MLLMs to effectively process and generate contextually relevant outputs by aligning and combining visual and textual information.

\subsubsection{Security Considerations in Multi-modal Systems}

The integration of multiple modalities in Multi-Modal Language Models (MLLMs) introduces unique security challenges. One significant concern is the emergence of cross-modal vulnerabilities, where adversarial inputs in one modality can compromise the system's integrity across other modalities. For instance, adversarial images can be crafted to manipulate the textual outputs of MLLMs, effectively bypassing safety mechanisms designed for text-only models \cite{shayegani2023jailbreak}. Alignment issues between modalities pose security risks. Discrepancies in the processing and interpretation of different modalities can be exploited to induce unintended behaviors in MLLMs. Attackers may leverage these misalignments to inject malicious content that is interpreted differently across modalities, leading to potential breaches in the system's defenses \cite{xu2024cross}.

The complexity of multi-modal systems increases the attack surface, providing more opportunities for adversaries to exploit. The integration points between modalities, such as the fusion of visual and textual data, can serve as potential attack vectors if not properly secured. Ensuring robust security in MLLMs requires addressing these cross-modal vulnerabilities and alignment challenges through comprehensive defense mechanisms and continuous evaluation of the system's resilience to adversarial attacks \cite{liu2024safety}.

Recent studies have highlighted the susceptibility of MLLMs to adversarial attacks that exploit these vulnerabilities. For example, Liu et al. demonstrated that MLLMs could be easily compromised by query-relevant images, as if the text query itself were malicious \cite{liu2024mm}. Similarly, Fan et al. discussed the potential perils of image inputs in MLLM security, emphasizing the need for robust defense mechanisms against such attacks \cite{fan2024unbridled}.

\subsection{Visual Processing Vulnerabilities}

\subsubsection{OCR Capabilities and Their Exploitation}

Optical Character Recognition (OCR) components in Multi-Modal Language Models (MLLMs) extract textual information from images, facilitating the integration of visual and textual data. However, these components are vulnerable to adversarial attacks that manipulate visual inputs to induce incorrect text extraction, compromising the model's integrity. For instance, adversarial perturbations can be applied to images, causing OCR systems to misinterpret or misread text, leading to erroneous downstream processing \cite{chen2020fawa}.

Recent studies have demonstrated the feasibility of such attacks. Chen et al. introduced the Fast Adversarial Watermark Attack (FAWA), which disguises perturbations as watermarks, making them less noticeable while achieving a high attack success rate against sequence-based OCR models \cite{chen2020fawa}. Similarly, Beerens and Higham analyzed the vulnerability of transformer-based OCR models, revealing that these systems are highly susceptible to both targeted and untargeted attacks, even with imperceptible perturbations \cite{beerens2023vulnerability}.

Xu et al. explored adversarial attacks on scene text recognition models, highlighting that both connectionist temporal classification (CTC) and attention-based models are susceptible to such attacks, resulting in significant performance degradation \cite{xu2020what}. Additionally, Shayegani et al. demonstrated that visual adversarial examples could bypass safety mechanisms in vision-language models, leading to the generation of harmful text outputs \cite{shayegani2023jailbreak, peng2024jailbreaking}.

\subsubsection{Cross-modal Training Weaknesses}

Inadequate cross-modal training can lead to misalignment between visual and textual representations in Multi-Modal Language Models (MLLMs), creating vulnerabilities that adversaries may exploit. Recent research has highlighted that such misalignments can be manipulated through adversarial attacks, where carefully crafted inputs in one modality (e.g., images) induce incorrect or harmful outputs in another (e.g., text) \cite{shayegani2023jailbreak, zhang2023adversarial}. For instance, Shayegani et al. demonstrated that adversarial images could bypass text-only alignment mechanisms in vision-language models, leading to the generation of undesirable content \cite{shayegani2023jailbreak}.

Furthermore, Leng et al. investigated hallucinations in large multimodal models, revealing that imbalances in modality integration and biases from training data contribute to these vulnerabilities \cite{leng2024curse}. Their findings underscore the necessity for balanced cross-modal learning to mitigate such issues. Additionally, Zhang et al. explored adversarial illusions in multi-modal embeddings, showing that attackers can manipulate inputs to align with arbitrary targets across modalities, thereby deceiving downstream tasks \cite{zhang2023adversarial}. These studies emphasize the importance of robust cross-modal training strategies to ensure proper alignment between modalities, thereby enhancing the security and reliability of MLLMs.

\subsubsection{Integration Points as Attack Surfaces}

In Multi-Modal Language Models (MLLMs), the junctures where visual and textual data converge are critical integration points that can be exploited as attack surfaces. Adversaries can introduce malicious inputs at these intersections, leading to compromised model performance. For instance, adversarial examples crafted in the visual domain can propagate through the model, resulting in incorrect textual outputs. This vulnerability is highlighted in studies where adversarial perturbations in images cause misclassifications in text recognition tasks \cite{chen2020fawa}.

Moreover, the alignment mechanisms between visual and textual modalities are susceptible to attacks. Research has demonstrated that manipulating the alignment process can induce the model to generate harmful or misleading content. For example, adversarial attacks targeting the alignment of image and text embeddings have been shown to mislead models into producing incorrect or harmful responses \cite{wang2024adashield}.

These findings underscore the necessity for robust defense strategies at the integration points of MLLMs. Enhancing the security of these junctures is essential to prevent adversarial exploitation and ensure the reliability of multi-modal systems.

\section{Visual Attack Vector Taxonomy}\label{sec:taxonomy}

\subsection{Adversarial Perturbation Attacks}

\subsubsection{Low-Cost Visual Manipulation}

Adversarial perturbation attacks subtly modify visual inputs to deceive models while maintaining high structural similarity index (SSIM) values to ensure imperceptibility. These attacks are computationally efficient and effective in misleading models. For example, the Fast Gradient Sign Method (FGSM) introduces minimal perturbations to images, causing significant misclassifications with minimal computational overhead \cite{goodfellow2015explaining}. Similarly, the Basic Iterative Method (BIM) extends FGSM by applying iterative perturbations, enhancing attack success rates while preserving visual similarity \cite{kurakin2017adversarial}.

Recent advancements have focused on optimizing perturbation directions to improve imperceptibility. Luo et al. proposed a frequency-driven approach that targets high-frequency components of images, resulting in perturbations that are less perceptible to the human eye while effectively deceiving models \cite{luo2022frequency}. Additionally, Chen et al. introduced a method leveraging diffusion models to generate imperceptible and transferable adversarial examples, demonstrating the potential for low-cost visual manipulations in complex scenarios \cite{chen2023diffusion}.

These methods highlight the ongoing development of efficient adversarial attacks that balance computational cost, effectiveness, and imperceptibility, posing significant challenges to the robustness of visual models.

\subsubsection{Cross-Modal Attack Integration}

Cross-modal attacks exploit the interplay between visual and textual modalities in multi-modal language models (MLLMs), crafting adversarial examples that simultaneously deceive both components, thereby amplifying the attack's effectiveness. Yin et al. introduced VLAttack, a method that generates adversarial samples by fusing perturbations of images and texts from both single-modal and multimodal levels, effectively disrupting universal representations and achieving high attack success rates across various tasks \cite{yin2023vlattack}. Similarly, Shayegani et al. developed compositional adversarial attacks on multi-modal language models, pairing adversarial images with textual prompts to break the alignment of the language model, highlighting the risk of cross-modality alignment vulnerabilities \cite{shayegani2023jailbreak}. Lu et al. proposed the Set-level Guidance Attack (SGA), which leverages cross-modal interactions and alignment-preserving augmentation to enhance adversarial transferability across different vision-language pre-training models \cite{lu2023set}.

\subsection{Advanced Attack Frameworks}

\subsubsection{VLATTACK Methodology}
VLATTACK is an adversarial framework that disrupts vision-language models by generating perturbations across both single-modal and multimodal levels. At the single-modal level, it applies a block-wise similarity attack (BSA) to create image perturbations that increase feature distance within network blocks, affecting downstream performance \cite{yin2023vlattack}. Text perturbations are generated independently through established text attack strategies, effectively targeting both visual and textual elements. When working under multi-modality, VLATTACK uses an iterative cross-search attack (ICSA), refining adversarial image-text pairs to enhance attack success across vision-language tasks. Experimental results show VLATTACK’s superior effectiveness over baseline methods.

\subsubsection{HADES Framework}

The HADES framework introduces a novel approach to compromising the alignment of multimodal large language models (MLLMs) by embedding harmful content within visual inputs, effectively executing visual jailbreaks. This method leverages the vulnerability of MLLMs to adversarial images, which can bypass content moderation systems and induce models to generate harmful outputs. The framework operates in three stages: first, it transfers harmful textual information into typographic images, redirecting the model's focus from text to visual content; second, it amplifies the harmfulness by attaching an image generated through prompt optimization; third, it integrates adversarial noise into the image to further disrupt the model's alignment. Experimental results demonstrate that HADES achieves high attack success rates against both open-source and proprietary MLLMs, highlighting the critical need for robust defense mechanisms against such adversarial attacks \cite{li2024images}.

\subsubsection{Sparse Adversarial Video Attacks}

Sparse Adversarial Video Attacks (SAVA) aim to deceive video recognition models by introducing minimal perturbations to select frames, thereby maintaining high perceptual quality. Mu et al. proposed DeepSAVA, a framework that combines additive perturbations with spatial transformations within a unified optimization framework. This approach uses the Structural Similarity Index Measure (SSIM) to quantify adversarial distance, ensuring that the perturbations remain imperceptible to human observers. The optimization process alternates between Bayesian optimization, which identifies the most influential frames, and stochastic gradient descent, which generates the perturbations. This method effectively balances attack success rates with the imperceptibility of the adversarial examples, highlighting vulnerabilities in current video recognition systems \cite{mu2021sparse}.

\subsubsection{Manifold-Aided Adversarial Examples}

Traditional adversarial attacks often rely on pixel-wise perturbations constrained by $\ell_p$ norms, which may result in limited attack transferability and reduced visual quality. To address these issues, Li et al. introduced a supervised semantic-transformation generative model that constructs an unrestricted adversarial manifold containing continuous semantic variations. This approach enables a legitimate transition from non-adversarial to adversarial examples, preserving semantic integrity and enhancing interpretability. Experiments on the MNIST and industrial defect datasets demonstrated that these manifold-aided adversarial examples exhibit superior visual quality and improved attack transferability compared to traditional methods, highlighting vulnerabilities in current deep neural networks \cite{li2024transcending}.

\subsubsection{AnyAttack Framework}

The AnyAttack framework introduces a self-supervised approach to generating targeted adversarial images for vision-language models (VLMs) without the need for label supervision. By utilizing a contrastive loss function, AnyAttack trains a generator on large-scale unlabeled image datasets, such as LAION-400M, enabling the creation of adversarial noise that effectively disrupts VLMs across various tasks, including image-text retrieval, multimodal classification, and image captioning. This method enhances the transferability of adversarial examples across different models, as demonstrated through extensive experiments on open-source VLMs like CLIP, BLIP, BLIP2, InstructBLIP, and MiniGPT-4, as well as commercial models such as Google's Gemini, Claude's Sonnet, and Microsoft's Copilot. The findings reveal significant vulnerabilities in current VLMs, underscoring the necessity for robust defense mechanisms \cite{zhang2024anyattack}.

\subsubsection{VT-Attack Methodology}

The VT-Attack methodology is designed to assess the robustness of large vision-language models (LVLMs) by generating adversarial images that disrupt the visual tokens produced by image encoders. This approach uses a multi-faceted strategy to perturb visual information at various levels. The feature attack component aims to deviate visual tokens from their original feature representations, thereby impairing the model's ability to accurately interpret visual data. Additionally, the relation attack focuses on altering the inherent relationships among visual tokens, further compromising the model's understanding of visual content. By integrating these strategies, VT-Attack effectively challenges the stability of the visual feature space within LVLMs, providing valuable insights into their vulnerabilities \cite{wang2024break}.

\subsubsection{InstructTA: Instruction-Tuned Targeted Attack}
InstructTA is a targeted gray-box adversarial attack framework for large vision-language models (LVLMs), where the adversary can access only the model’s visual encoder, without prompts or language model (LLM) knowledge \cite{wang2023instructta}. Using a public text-to-image model, InstructTA generates a target image linked to a desired response, while GPT-4 infers an instruction aligning with this response. Together, these create a local surrogate model that mimics the victim LVLM's visual encoder, optimizing adversarial examples by minimizing feature distance between target and adversarial images. To enhance cross-prompt transferability, InstructTA uses instruction paraphrasing, allowing it to maintain high attack success rates across models and prompts, exposing critical vulnerabilities in LVLMs.

\subsubsection{PG-Attack: Precision-Guided Adversarial Attack Framework}

The PG-Attack framework introduces a novel approach to generating adversarial examples targeting vision-language models, particularly in autonomous driving contexts. It integrates two primary techniques: Precision Mask Perturbation Attack (PMP-Attack) and Deceptive Text Patch Attack (DTP-Attack). PMP-Attack focuses on precisely targeting specific regions within an image to minimize overall perturbation while maximizing the impact on the model's feature representation of the target object. DTP-Attack introduces deceptive text patches that disrupt the model's scene understanding, further enhancing the attack's effectiveness. This combined approach effectively deceives advanced multimodal models, including GPT-4V, Qwen-VL, and imp-V1, highlighting vulnerabilities in current vision-language systems \cite{fu2024pg}.

\subsubsection{Efficient and Effective Universal Adversarial Attack}

The Direct Optimization-based Universal Adversarial Perturbation (DO-UAP) method presents an efficient way when generating universal adversarial perturbations (UAPs) targeting vision-language pre-training (VLP) models. Traditional generator-based UAP methods are often resource-intensive and time-consuming, while DO-UAP uses a direct optimization strategy that significantly reduces computational overhead while maintaining high attack efficacy. This method incorporates a multimodal loss design, ensuring that the perturbations effectively disrupt both visual and textual components of VLP models. Additionally, a data augmentation strategy is introduced to enhance the transferability of the adversarial perturbations across different models and tasks. Extensive experiments demonstrate that DO-UAP achieves superior attack performance with a 23-fold reduction in time consumption compared to existing methods, highlighting its potential for practical applications in evaluating and improving the robustness of VLP models \cite{yang2024efficient}.

\subsubsection{Collaborative Multimodal Adversarial Attack (Co-Attack)}

The Collaborative Multimodal Adversarial Attack (Co-Attack) is a method designed to enhance the effectiveness of adversarial attacks on vision-language pre-training (VLP) models by simultaneously perturbing both image and text modalities. Traditional adversarial attacks often focus on a single modality, which may not fully exploit the vulnerabilities inherent in multimodal models. Co-Attack addresses this by introducing coordinated perturbations across modalities, thereby increasing the likelihood of misleading the model's predictions. This approach has been evaluated on various VLP architectures and downstream tasks, demonstrating improved attack success rates compared to unimodal attacks \cite{zhang2022towards}.

\section{Attack Characteristics and Implications} \label{sec:analysis}

\subsection{Performance Analysis}

\subsubsection{Success Rates Across Different Models}

Evaluating adversarial attack success rates across various vision-language models reveals differing vulnerabilities, underscoring the necessity for tailored defenses. For instance, the VLATTACK framework demonstrated high attack success rates on multiple tasks, including visual question answering and visual reasoning, when targeting models such as BLIP, CLIP, and OFA \cite{yin2023vlattack}. Similarly, the Joint Multimodal Transformer Feature Attack (JMTFA) achieved significant success against vision-language pretrained transformers, indicating that model size does not correlate with adversarial robustness \cite{guan2024probing}. Furthermore, studies have shown that even advanced models like GPT-4V are susceptible to imperceptible adversarial perturbations, leading to incorrect image captioning outputs \cite{schlarmann2023adversarial}. These findings highlight the varying degrees of susceptibility among models and emphasize the importance of developing customized defense mechanisms to enhance adversarial robustness.

\subsubsection{Visual Similarity Preservation}

When working on adversarial attacks on vision-language models, maintaining high visual fidelity ensures that perturbations remain inconspicuous to humans while deceiving the model. Metrics such as the Structural Similarity Index Measure (SSIM) are commonly used to quantify similarity, focusing on luminance, contrast, and structure \cite{wang2004image}. Recent approaches like VT-Attack and Sim-CLIP have focused on achieving imperceptibility; VT-Attack disrupts visual tokens from image encoders, deceiving vision-language models without altering image appearance, while Sim-CLIP fine-tunes models for enhanced robustness against adversarial attacks \cite{wang2024break} \cite{hossain2024sim}. The success of such methods emphasizes the challenge posed by imperceptible perturbations in maintaining the security of vision-language models.

\subsubsection{Attack Transferability Patterns}

Analyzing adversarial transferability between models is crucial for understanding cross-model vulnerabilities and enhancing defense strategies. Recent research on vision-language models (VLMs) has highlighted significant generalization patterns in adversarial attacks. Lu et al. introduced the Set-level Guidance Attack (SGA), which improves transferability by utilizing cross-modal interactions and alignment-preserving augmentations, demonstrating that SGA-generated adversarial examples could effectively exploit vulnerabilities across different VLMs \cite{lu2023set}. Han et al. proposed OT-Attack, employing optimal transport optimization to align feature distributions of adversarial and clean samples, achieving higher success rates in black-box settings and showing that certain feature alignments are more prone to attack transferability \cite{han2023ot}. 

\subsection{Security Implications}

Adversarial attacks on vision-language models (VLMs) present a universal threat by bypassing model safeguards, which significantly compromises model alignment and enables unauthorized actions. Recent research demonstrates that universal adversarial perturbations (UAPs) can effectively mislead VLMs across a range of tasks, showing a high degree of transferability. Zhang et al. introduced a method to generate UAPs capable of deceiving multiple VLMs in diverse downstream tasks, emphasizing the broad applicability of such attacks \cite{zhang2024universal}. Similarly, Shayegani et al. have developed compositional adversarial attacks that combine adversarially manipulated images with benign textual prompts to disrupt the alignment of VLMs, thereby triggering harmful text generations, even when attackers only have access to the vision encoder \cite{shayegani2023jailbreak}. This approach reveals the susceptibility of models as well as demonstrating how relatively minimal access requirements can pose significant risks. A related study showed that universal and transferable adversarial attacks can bypass safety mechanisms in aligned language models, showing the feasibility of these attacks across different architectures and transferability \cite{zou2023universal}.

Adversarial attacks on vision-language models (VLMs) present significant privacy and safety risks by manipulating models to generate harmful outputs or extracting sensitive training data, which compromises user trust \cite{qi2023visual, carlini2021extracting, peng2024securing}. These vulnerabilities necessitate robust defense mechanisms and thorough evaluations to ensure privacy and safety in AI systems.

\section{Defense Mechanisms} \label{sec:defense}

\subsection{Current Approaches}

\subsubsection{SmoothVLM Framework Implementation}

The SmoothVLM framework enhances the robustness of vision-language models (VLMs) against adversarial patch attacks by employing randomized smoothing techniques. Adversarial patches are malicious modifications to input images designed to mislead VLMs into generating incorrect or harmful outputs. SmoothVLM mitigates this threat by introducing pixel-wise random perturbations to the input images, effectively neutralizing the adversarial impact of such patches. This approach significantly reduces the attack success rate to between 0\% and 5\% on leading VLMs, while maintaining approximately 67.3\% to 95.0\% context recovery of benign images, thereby preserving model performance \cite{sun2024safeguarding}.

\subsubsection{Pixel-wise Randomization Techniques}
Implementing pixel-wise randomization introduces controlled noise to input images, disrupting adversarial perturbations and improving model resilience against visual attacks \cite{park2022randomize}. 

\subsubsection{Context-Based Resilience Methods}

Context-based resilience methods enhance the robustness of multimodal language models by leveraging semantic understanding to detect and mitigate adversarial inputs. These strategies analyze the contextual coherence between different modalities, such as text and images, to identify inconsistencies indicative of adversarial attacks. By integrating semantic analysis, models can adaptively recognize and counteract sophisticated threats, thereby improving their resilience. For instance, the CosPGD attack framework introduces an efficient white-box adversarial attack for pixel-wise prediction tasks, emphasizing the importance of context in developing robust defense mechanisms \cite{agnihotri2023cospgd}. Implementing such context-aware defenses is crucial for maintaining the integrity and reliability of multimodal language models in diverse applications.

\subsubsection{MirrorCheck Defense Mechanism}

MirrorCheck is an adversarial defense mechanism designed to enhance the robustness of vision-language models (VLMs) against adversarial attacks. This approach utilizes text-to-image (T2I) models to generate images from captions produced by the target VLM. By comparing the embeddings of the original and generated images in the feature space, MirrorCheck identifies discrepancies indicative of adversarial manipulation. Empirical evaluations demonstrate that MirrorCheck outperforms baseline methods adapted from image classification domains, effectively detecting adversarial samples and improving the security of VLMs \cite{fares2024mirrorcheck}.

\subsection{Future Directions}

\subsubsection{Robust Defense Mechanism Development}

Research on robust defense mechanisms for multi-modal language models shall focus on adaptive methods that proactively counter emerging adversarial tactics. Enhancing visual encoders to ensure resilience, potentially using advanced training techniques that reinforce the similarity between perturbed and original inputs, is a promising approach. Researchers could also develop dual architectures for consistency against perturbations. Another key direction is the creation of detection systems that identify adversarial inputs through cross-verification or generative comparisons to catch discrepancies effectively. Additionally, incorporating randomization and transformation techniques could mitigate adversarial attacks while retaining benign context. These defenses must be both lightweight and effective, balancing minimal computational overhead with maximum model security, thus ensuring the robustness and reliability of multi-modal models against evolving threats.

\subsubsection{Evaluation Metric Improvements}

Evaluation metrics for MLLMs would focus on establishing more sophisticated frameworks that can holistically assess model robustness against adversarial attacks. Traditional metrics often lack the depth needed to capture the intricate dynamics of multi-modal interactions, suggesting a need for comprehensive and specialized benchmarks. As vision-language models become increasingly integrated into sensitive applications, future evaluations must address not only performance but also security aspects, identifying vulnerabilities across different modalities. For instance, recent frameworks designed to measure model understanding and reasoning highlight the necessity of developing metrics that extend beyond conventional accuracy assessments \cite{lee2024vhelm}. Additionally, recent benchmarks introduced for MLLMs aim to evaluate both perceptual and cognitive tasks, serving as a foundation for more thorough and security-aware performance assessments \cite{fu2023mme}. Future metrics must encompass a broader range of factors, including robustness under adversarial conditions, response consistency across modalities, and adaptability in dynamic, real-world environments. As suggested by recent surveys, advancing evaluation methodologies will be essential in guiding the development of next-generation MLLMs that are not only highly accurate but also resilient to exploitation, enhancing their reliability and trustworthiness in practical deployment \cite{li2024survey}.

\subsubsection{Cross-modal Security Integration}

Integrating security measures across modalities is essential for developing unified protection schemes that address vulnerabilities at the intersection of visual and textual data in multi-modal systems. Recent research highlights the necessity of cross-modal defense mechanisms to safeguard these systems against adversarial attacks. For instance, a study by Gong et al. introduced a cross-modality perturbation synergy attack for person re-identification, emphasizing the need for integrated security strategies to counteract such threats \cite{gong2024cross}. Similarly, Yang et al. proposed a security matrix for multimodal agents on mobile devices, underscoring the importance of systematic security assessments and defenses in multi-modal contexts \cite{yang2024security}. These findings suggest that future research should focus on developing comprehensive security frameworks that encompass all modalities, ensuring robust protection against complex, cross-modal adversarial attacks.

\section{Conclusion and Future Work} \label{sec:conclusion}

There are several significant challenges when defending MLLMs against visual attacks:

1. \textbf{Complexity of Multi-Modal Integration:} MLLMs process and integrate information from multiple modalities, such as text and images. This integration increases the attack surface, making it difficult to design defenses that effectively address vulnerabilities across all modalities \cite{pi2024mllm, peng2024securing}.

2. \textbf{Adversarial Robustness:} MLLMs are susceptible to adversarial examples—inputs intentionally designed to cause the model to make errors. Crafting robust defenses against such attacks is challenging due to the high dimensionality and continuous nature of visual data \cite{cui2023robustness}.

3. \textbf{Evaluation Metrics:} Existing evaluation metrics often fail to capture the complexities of multi-modal interactions, making it difficult to assess the effectiveness of defense mechanisms accurately. Developing comprehensive evaluation frameworks tailored to MLLMs is essential for meaningful assessments \cite{fu2023mme}.

4. \textbf{Transferability of Attacks:} Adversarial attacks can transfer across different models and tasks, complicating the development of defenses. A defense mechanism effective against one type of attack may not generalize well to others, necessitating adaptable and comprehensive defense strategies \cite{zhou2024adversarial}.

5. \textbf{Computational Constraints:} Implementing robust defense mechanisms often requires significant computational resources, which may not be feasible in real-time applications. Balancing security and efficiency remains a critical challenge in deploying MLLMs in practical scenarios \cite{pi2024mllm}.

Overcoming these challenges requires more complete methods that consider MLLM characteristics and evolving nature of adversarial threats.

Research on visual attacks in MLLMs enhances both the understanding and robustness of these systems. A primary focus lies in cross-modal adversarial attacks, which exploit the intricate dependencies between visual and textual modalities and uncover distinct vulnerabilities within. Universal adversarial perturbations—generalizable attacks reveal the model's systemic weaknesses, offering insights into model-agnostic flaws. Creating sophisticated benchmarking frameworks that accurately capture and evaluate cross-modal interactions establish a standardized foundation for assessing adversarial robustness. Developing targeted defense mechanisms, such as adversarial training protocols and architectural safeguards, increases the model's resilience against visual attack vectors. Investigating the transferability of adversarial attacks across models and tasks leads a pathway to scalable, generalized defense strategies that can adapt to varied architectures. These methods are indispensable to building resilient MLLMs capable of handling evolving adversarial attacks and extending their utility in complex, real-world applications.

\bibliographystyle{ieeetr}  
\bibliography{references} 


\end{document}